\documentclass[9pt,journal,compsoc]{IEEEtran}
\ifCLASSOPTIONcompsoc
  \usepackage[nocompress]{cite}
\else
  \usepackage{cite}
\fi
\ifCLASSINFOpdf
\else
\fi
\usepackage{soul, physics, graphicx, xcolor}
\usepackage{dblfloatfix, makecell}
\usepackage{enumitem}
\setitemize{noitemsep,topsep=0pt,parsep=0pt,partopsep=0pt}

\begin{document}
\title{Approximate Quantum Random Access Memory Architectures}

\author{Koustubh Phalak,~\IEEEmembership{Student Member,~IEEE,}
        Junde Li,~\IEEEmembership{Student Member,~IEEE,}
        Swaroop Ghosh,~\IEEEmembership{Senior Member,~IEEE}% <-this % stops a space
}


\markboth{}%
{Shell \MakeLowercase{\textit{et al.}}: Approximate Quantum Random Access Memory for Machine Learning and Storage}



\IEEEtitleabstractindextext{%
\begin{abstract}
Quantum supremacy in many applications using well-known quantum algorithms rely on availability of data in quantum format. Quantum Random Access Memory (QRAM), an equivalent of classical Random Access Memory (RAM), fulfills this requirement. However, the existing QRAM proposals either require qutrit technology and/or incur access challenges. 
%For example, writing of the data is performed efficiently while reading remains a difficult task. 
We propose an approximate Parametric Quantum Circuit (PQC) based QRAM which takes address lines as input and gives out the corresponding data in these address lines as the output. We present two applications of the proposed PQC-based QRAM namely, storage of binary data and storage of machine learning (ML) dataset for classification. For the machine learning task, we perform binary classification using three different setups, (a) QNN with QRAM (QRAM + QNN), (b) QNN with normal data embedding and (c) Fully Connected Neural Network (FCNN). We show that QRAM + QNN converges (with 100\% classification accuracy) faster i.e., by $6^{th}$ epoch, compared to the other two setups which converge at around $10^{th}$ and $15^{th}$ epochs. The loss for QRAM + QNN (0.53) is less than FCNN setup (0.89) and comparable to QNN with normal embedding (0.48). For the storage of binary data, 
%we present two different setups, one without clustering and one with agglomerative clustering. 
we evaluate Hamming Distance (HD) and percentage correct prediction metrics to quantify the performance. 
%of the binary data QRAM. 
We observe an increase in HD from 0 to 3.97 and decrease in percentage of correct predictions from 100$\%$ to 0.58$\%$ as we increase the number of address and data lines from 2 to 9. We propose agglomerative clustering of data as a pre-processing step before training the QRAM to improve the HD i.e., from 3.97 to 2.17 and increase the percentage correct predictions to 10.1\% from 0.58\% till 9 address lines.
%\hl{fix--(from 0 to 2.17) and reduce percentage correct predictions (from 100\% to 10.1\%)}. 
%We also show that for all address lines, performing clustering pre-processing yields more efficient results as compared to the case without clustering.  

% We show 100\% classification accuracy for training and testing for binary classification for 9 address lines. QRAM + quantum neural network (QNN) converges faster than QNN with data embedding and Fully Connected Neural Network (FCNN) (around 5 epochs), and lower final loss (0.53) than FCNN (0.89). For binary data QRAM, we observe an increase in Hamming distance from 0 to 3.97 and decrease in percentage of correct predictions from 100$\%$ to 0.58$\%$ as we increase the number of address and data lines from 2 to 9.
%For the binary data QRAMs, we show that the QRAM works with zero hamming distance only till 2 address lines and shows non-zero hamming distance for higher number of address lines. 
\end{abstract}

}

    \maketitle

\IEEEdisplaynontitleabstractindextext

% \fi
\IEEEpeerreviewmaketitle

\IEEEraisesectionheading{\section{Introduction}\label{sec:introduction}}

%\begin{figure*}[t]
    %\centering
    %\includegraphics[width=\linewidth]{fig/Existing_QRAMs.pdf}
 %   \caption{Existing QRAM architectures. \hl{(a)Bucket Brigade QRAM (Qutrit-based) (b)Circuit-based QRAM (c)EQGAN QRAM. }}
%    \label{fig:existing_qrams}
%\end{figure*}

\IEEEPARstart{R}{ecent} advances of quantum computing in various fields such as, machine learning (ML) have shown great potential. Machine learning methods augmented with quantum computing like quantum clustering, quantum decision trees, quantum support vector machines, and quantum neural networks (QNNs) \cite{schuld2015introduction} are able to provide quantum speedup using quantum algorithms compared to their classical counterparts \cite{biamonte2017quantum}. An important aspect for such augmented quantum algorithms is the conversion of classical data into quantum domain. Encoding methods like basis embedding, amplitude embedding, angle embedding \cite{schuld2018supervised} (Chapters 5 and 6) use various mathematical formulae to embed classical data onto qubits. Basis embedding encodes binary data of $n$ bits onto $n$ qubits. Amplitude embedding normalizes $2^n$ data and embeds them onto $n$ qubits. Angle embedding performs rotation operation on qubits, either using RX, RY or RZ gates and embeds $n$ classical datapoints onto $n$ qubits. While these methods are useful for small and simple datasets, they are not efficient for more complex datasets. Quantum Random Access Memory (QRAM) to store and load quantum data could be much more effective for quantum machine learning problems. The user should be able to store new classical data or update it onto the QRAM and load the stored quantum data as needed.
% The user should be able to store new classical data or update it onto the QRAM and load the stored quantum data for a general application as needed. 
% To formally define, a QRAM is a component that just like its classical equivalent, takes in address as input and gives output as the data present in the corresponding address. The QRAM has \emph{write} operation to store the new data, and \emph{read} operation to load the stored data.

In this paper, we propose a Parametric Quantum Circuit (PQC)-based trainable approximate QRAM which takes address as input and gives output as the data for that address. We say that the QRAM is approximate because the predicted output of QRAM will not be exactly same as the input. This happens because data loading methods for NISQ era quantum computers are not perfect and there are different errors in execution of quantum circuits. We introduce relevant background on quantum computing and quantum machine learning and previously published related works on QRAM in Section 2 and the architectures for the proposed QRAM in Section 3. We then show two applications for the proposed QRAM architecture: one for storage of digit images for performing binary classification and other for storage of binary data, in Sections 4 and 5 respectively. We give concluding remarks in Section 6.
%We also show two applications for the proposed QRAM structure for binary classification of data and storage of binary data in Sections 4 and 5 respectively.
\emph{ To the best of knowledge, this is the first approximate QRAM architecture which can be used to load arbitrary data in quantum form to any quantum circuit.}
\vspace{-0.25cm}

\section{Background and Related Works}
\subsection{Background}
\textbf{\textit{Qubits:}} Qubits are the fundamental units of a quantum computer (analogous to classical bits for a classical computer). Unlike classical bits, qubits store information in the form of states represented using ket notation $\ket{\psi}=$
$\big[\begin{smallmatrix}
  a\\
  b
\end{smallmatrix}\big]$ where $a^2$ denotes the probability of the qubit being measured to 0 and $b^2$ denotes the probability of the qubit being measured to 1. Each qubit has two basis states, one with $a=1, b=0$ (defined as $\ket{0}$) and one with $a=0,b=1$ (defined as $\ket{1}$).
There are different types of qubits like superconducting qubits, ion trap qubits, photonic qubits, neutral atom qubits, etc. \vspace{0.05cm}

\noindent \textbf{\textit{Quantum Gates:}} A quantum gate is a unitary operation performed on a qubit that changes the state of the said qubit. Every quantum gate can be represented as a unitary matrix, and can work on either a single qubit or more than one qubits. The most common ones are single qubit gates like hadamard gate (H), bit-flip gate (X), Pauli Y and Z gates, and two qubit gates like CNOT gate, SWAP gate, and controlled Pauli gates. There are also special gates like reset gate and measurement gate.
% There are also special gates like reset gate that resets the state of a qubit to $\ket{0}$ and measurement gate that measures the classical output for a particular qubit.
% \vspace{0.05cm}

\noindent \textbf{\textit{Quantum Circuit:}} A quantum circuit is a program that contains an ordered sequence of various quantum gates that act on specified number of qubits and may contain intermediate reset and measurement operations. A user typically builds a quantum circuit, and obtains classical measurement output back. Based on the gates used and their placement, various quantum circuits %(such as, Quantum Approximate Optimization Algorithm (QAOA), Shor's Algorithm and Grover's Algorithm)
can be used for targeted applications. %  are some famous examples of such quantum circuits.
\vspace{0.05cm}

\noindent \textbf{\textit{Parametric Quantum Circuit:}} A Parametric Quantum Circuit (PQC) is a special type of quantum circuit where user can feed custom tunable parameters to change the output of the circuit according to the value of the parameters fed (typically encoded as rotation angle of a single qubit gate). It is a trainable quantum circuit which can be trained like a normal ML model.\vspace{0.05cm}

\noindent \textbf{\textit{QNN:}} Quantum Neural Network (QNN) is a trainable PQC that has hidden layers similar to a classical neural network. All the layers have trainable parameters that are updated every epoch.
% All the layers have trainable parameters that are updated after every iteration of training.  

% \paragraph{\textit{Hamming Distance:}} Hamming distance is a metric used to measure how different are binary data from each other. For two binary strings $b_1$ and $b_2$, the hamming distance between $b_1$ and $b_2$ is defined as the number of bit positions where the bit value is different. In the context of QRAM, we would be using this measure as an evaluation metric to test the performance of binary data QRAM. $b_1$ will be the predicted bit string data, and $b_2$ will be the actual bit string data. Based on $b_1$ and $b_2$ values, we would evaluate the hamming distance. Ideally, for a perfectly accurate binary data QRAM, the average hamming distance for all data would be zero. However, since the QRAM proposed is an approximate QRAM, there will be errors. We later propose a clustering method to mitigate the error in binary data QRAM.

\begin{table*}[t]
    \centering
    \includegraphics[width=0.95\linewidth]{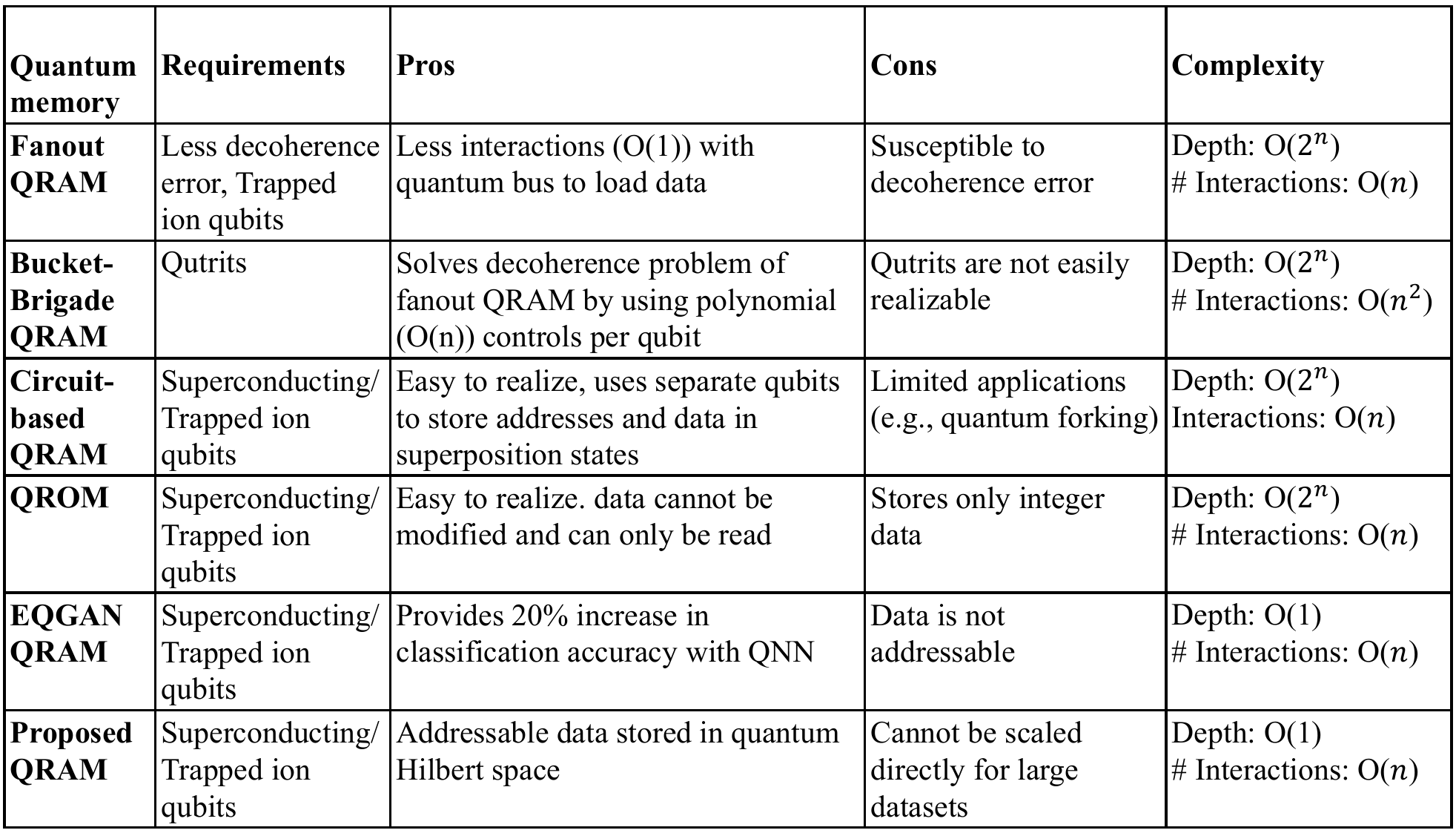}
    \caption{Comparison of various quantum memories.}
    \label{tab:related_works_comparison}
    \vspace{-8mm}
\end{table*}

\begin{figure}[t]
    \centering
    \includegraphics[width=0.85\linewidth]{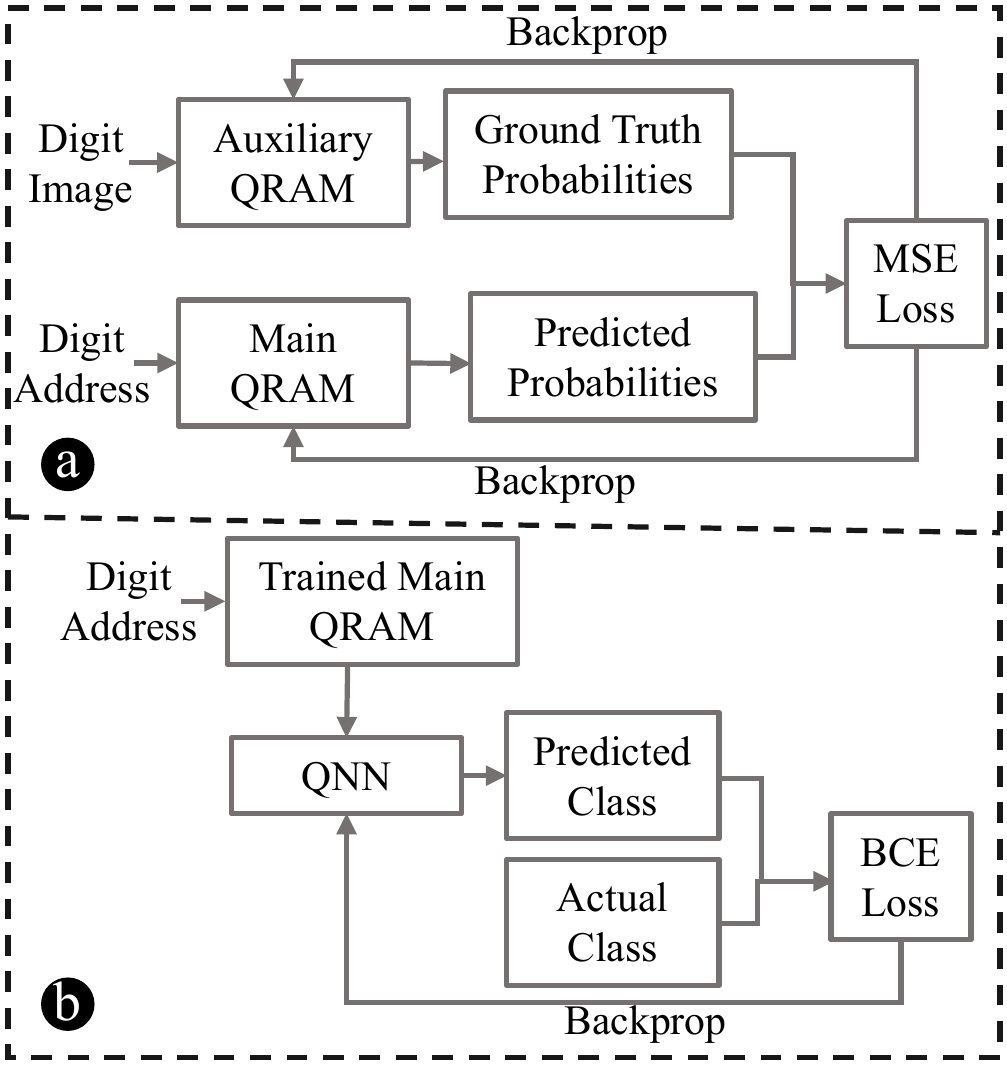}
    \caption{Classification using QRAM + QNN setup. (a) Training of QRAM with auxiliary and main PQCs and (b) training of QNN with using trained QRAM for classification.}
    \label{fig:qram_ml_task}
    \vspace{-4mm}
\end{figure}

% \paragraph{\textit{Agglomerative Clustering:}} Agglomerative Clustering is a bottom-up approach to cluster data where each datapoint is considered to be a single cluster at the start of the clustering algorithm, and the datapoints are gradually merged in each iteration into a single cluster. Based on the number of clusters provided at the start of the algorithm, agglomerative clustering finally groups the data into those many number of groups. In the binary data QRAM, we propose an agglomerative clustering pre-processing on the data prior to sending it to the QRAM in order to optimize the hamming distance and percentage correct predictions.

\subsection{Related Works}
Various implementations of QRAM have been proposed in literature previously like bucket brigade QRAM \cite{giovannetti2008quantum}, circuit-based QRAM \cite{park2019circuit} and entangling quantum generative adversarial network \cite{niu2022entangling}. These implementations, however, pose a few problems with respect to applicability of QRAM for quantum machine learning tasks. \cite{giovannetti2008quantum,park2019circuit,niu2022entangling} all use QRAM that write and read data into superposition. The work presented in \cite{giovannetti2008quantum} relies on qutrit technology. However, so far such technology is not realizable easily. \cite{park2019circuit} is easy to replicate since they used a quantum circuit-based QRAM, however the read operation they use is not well-defined, and moreover the data that can be stored can only be integer. The authors show an application of the stored superposition QRAM values for quantum forking. However the extent of its utility is limited \cite{park2019parallel}. The authors in \cite{niu2022entangling} preprocess data in quantum using EQGAN (Entangling Quantum Generative Adversarial Network) which is a circuit-based quantum GAN that is used as a QRAM to store data. They show better results with EQGAN for quantum classification application. However, EQGAN does not have addressable data like a classical RAM. The above challenges makes it difficult to realize a practical QRAM which can be used for various applications. 

Other works have also been proposed that discuss QRAM for optimization of quantum machine learning (QML) tasks (\cite{gyongyosi2020optimizing}).
%bang2019optimal, ramezani2020machine, adcock2015advances
Modification and efficient implementations to overcome challenges of the original QRAMs are also proposed. For example, 
%\cite{hann2019hardware} and 
\cite{hong2012robust} present efficient implementation of the bucket-brigade QRAM \cite{giovannetti2008quantum} whereas \cite{de2020circuit} and \cite{park2019parallel} are extensions of the circuit-based QRAM \cite{park2019circuit}. A comparison of various quantum memories is shown in Table \ref{tab:related_works_comparison}.
\vspace{-7mm}

\section{Proposed QRAM architecture}
\subsection{Basic architecture}
\begin{figure}[t]
    \centering
    \includegraphics[width=0.8\linewidth]{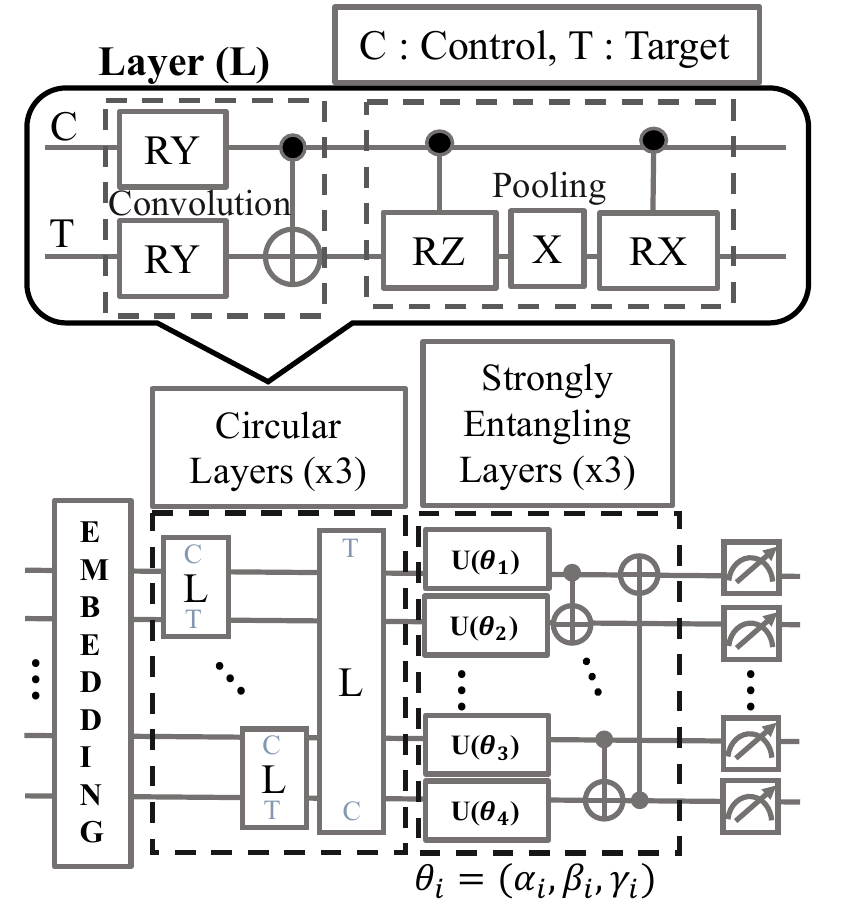}
    \vspace{-3mm}
    \caption{Basic QRAM Architecture. Strongly entangling layers are replicated from \cite{stronglayers} which in turn are inspired from \cite{schuld2020circuit}.}
    \label{fig:basic_qram_arch}
    \vspace{-4mm}
\end{figure}

The proposed QRAM (Fig. \ref{fig:basic_qram_arch}) is a trainable PQC, which has $n$ address lines for less than $2^n$ datapoints. For each individual datapoint, a unique address is provided, and an address-data pair is formed. We use the convolution ansatz 1 and pooling layer from \cite{hur2022quantum} as the basic layer for the proposed QRAM although other circuits are also possible. The reason for choosing this particular ansatz is that it has the lowest circuit depth and gate count. This will reduce the overall depth of the QRAM circuit which in turn can provide better fidelity of computation on real quantum hardware. We combine the ansatz in a circular fashion followed by strongly entangling layers to provide strong entanglement between qubits. According to \cite{cai2015entanglement}, having strongly entangled qubits can provide exponential speedup for classification problems.
% \color{blue} The reason for choosing this particular ansatz is that the said ansatz has the lowest circuit depth and gate count. This is will reduce the overall depth of the QRAM circuit which in turn can provide better fidelity of computation on real quantum hardware. We combine the ansatz in a circular fashion followed by strongly entangling layers to provide strong entanglement between qubits. According to \cite{cai2015entanglement}, having strongly entangled dataset can provide exponential speedup for classification problems. \color{black}
%The convolution ansatz 1 consists of 2 RY gates on two qubits and one CNOT gate between the 2 qubits, and the pooling layer consists of a sequence of controlled RZ gate, Pauli X gate and controlled RX gate (Fig. \ref{fig:basic_qram_arch} inset). 
We first load the address into qubits using any $n$ bits to $n$ qubits embedding like either angle embedding or basis embedding, and then pass the embedded quantum data through a sequence of the layers defined in Fig. \ref{fig:basic_qram_arch}. These layers are placed in circular fashion. Finally, three strongly entangling layers \cite{stronglayers}, (which is inspired from \cite{schuld2020circuit}) are placed at the end before taking measurement on all qubits. We note that the proposed QRAM is an application of PQC that can be used to load data. This PQC is called a QRAM since it provides addressability to data like classical RAMs and stores data in quantum form. %This basic architecture is shown in Fig. \ref{fig:basic_qram_arch}.
\vspace{-4mm}

\subsection{Training architecture of QRAM}
We use the basic QRAM architecture defined above to design QRAMs for each application. For the ML task, we use an auxiliary QRAM which takes digit image and generates ground truth probabilities for the qubits and the main QRAM which takes address of the digit image and outputs predicted probabilities (Fig. \ref{fig:qram_ml_task}(a)). Training of QRAM is done in two steps; in the first step both the auxiliary QRAM and main QRAM parameters are trained so that the ground truth probabilites and the predicted probabilities are as close to each other as possible, and in the second step the  parameters of the auxiliary QRAM are fixed and only the main QRAM parameters are trained to further reduce the loss. Both steps involve training the QRAM for 100 epochs each. After training, the auxiliary QRAM is discarded and the output of the main QRAM for each address is provided to a QNN in quantum format for classification (Fig. \ref{fig:qram_ml_task}(b)). For the binary data QRAM, a single basic architecture PQC is used which outputs Pauli-Z expectation value measurement to output individual data bits on all qubits. For both the scenarios, the proposed QRAM is approximate since it predicts the output in each address approximately rather than accurately. For ML QRAM, the approximation error is difference between the predicted probabilities and ground truth probabilities for every image, and for binary data QRAM, the approximation error arises from the Hamming distance between predicted data bits and actual data bits. Note that the proposed QRAM architecture can handle only sequential access of data and does not scale for datasets of larger sizes for which we can use multiple parallel banks of QRAMs.
% Note that the proposed QRAM architecture can handle only sequential access of data and does not scale for datasets of larger sizes. For larger datasets, we can use multiple banks of QRAMs to store the data in quantum format. \color{black} 
% you can discuss these in respective sections--We propose two types of QRAM architectures, one for storing integral data in binary data, and another for classifying digit images. For each QRAM, we use the basic architecture described above and shown in Fig. \ref{fig:basic_qram_arch}. For the binary data QRAM, we measure Pauli-Z expectation value and for the digit images QRAM, we use a system of two basic architecture QRAMs (auxiliary for digit images, main for address bits) and give probability output of all states.
\vspace{-0.4cm}

% \vspace{-2mm}

\section{QRAM for machine learning tasks}
\subsection{Experimental Setup}
We augment a QRAM with a QNN to perform binary classification of 0 and 1 digits in digits dataset provided by UCI Machine Learning Repository. We train the QRAM to load the digit images, and compare the performance with quantum classifier with embedding and classical Fully Connected Neural Network (FCNN).
%We perform binary classification using three different setups. In the first setup, we use the alternate QRAM architecture along with a QNN augmented. The second setup will only have a QNN which directly takes input as the images and performs the classification. In the third scenario, we use only a classical Fully Connected Neural Network (FCNN) to perform the binary classification. For all the three setups, 
We report the training and testing losses, training and testing accuracies for the three cases. 
%We also provide a comparison between the results of all three cases and interpretation of the results. 
For all the setups, we maintain a batch size of 16 and learning rate of 0.001. Since the number of images of both classes combined are 360 , we use 9 address lines to fit all images. Both the QRAM and the QNN use the same basic architecture as described above. We also performed noisy simulations and experimental inferencing on a shorter dataset of 0 and 1 classes of 64 images with noise characteristics of IBM Jakarta. We train this dataset on ideal simulator for 100 epochs and noisy simulator for 100 epochs.
\vspace{-0.4cm}

\subsection{Classification Results}
\begin{figure*}[t]
    \centering
    \includegraphics[width=0.9\linewidth]{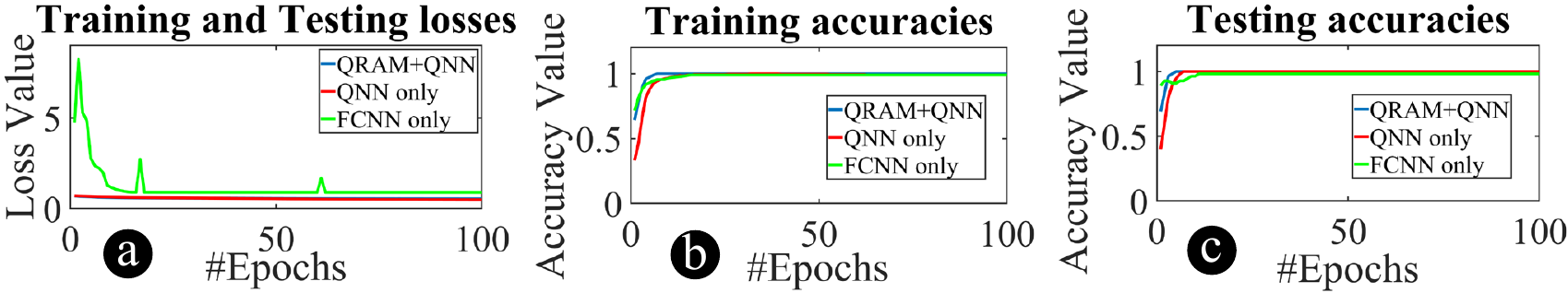}
    \vspace{-3mm}
    \caption{Experimental results for binary classification task for all setups, (a) training and testing losses, (b) training accuracies, and (c) testing accuracies}
    \label{fig:ml_experimental_results}
    \vspace{-6mm}
\end{figure*}

\
%The classification results for all the three experimental setups are shown in Fig. \ref{fig:ml_experimental_results}. 
%In , we report the training and testing losses. 
For each case, the training and testing loss curves (Fig. \ref{fig:ml_experimental_results}(a)) are nearly identical. This is because the logarithmic factor in binary cross entropy brings the training and testing loss values very close to each other. Fig. \ref{fig:ml_experimental_results}(b) and (c) reports the training and testing accuracies respectively. We observe 100\% classification accuracy for all the three setups. We note that the QRAM + QNN and QNN with embedding only setups outperform FCNN in terms of loss and convergence. QRAM + QNN converges the fastest at around $6^{th}$ epoch, QNN with embedding converges around $10^{th}$ epoch, and FCNN converges at around $15^{th}$ epoch. The final loss for QRAM + QNN is around 0.53, for QNN with embedding is 0.48 and for FCNN only is 0.89. The proposed QRAM + QNN setup also outperforms the EQGAN performance (\cite{niu2022entangling}) which achieves around 65\% classification accuracy on a dataset of two classes of normal distributions. From the noisy simulation results, we observe 100\% testing accuracy on Pennylane and IBM QASM simulators and roughly 63\% testing accuracy for IBM Jakarta and IBM Oslo hardware.

\begin{figure}[t]
    \centering
    \includegraphics[width=0.9\linewidth]{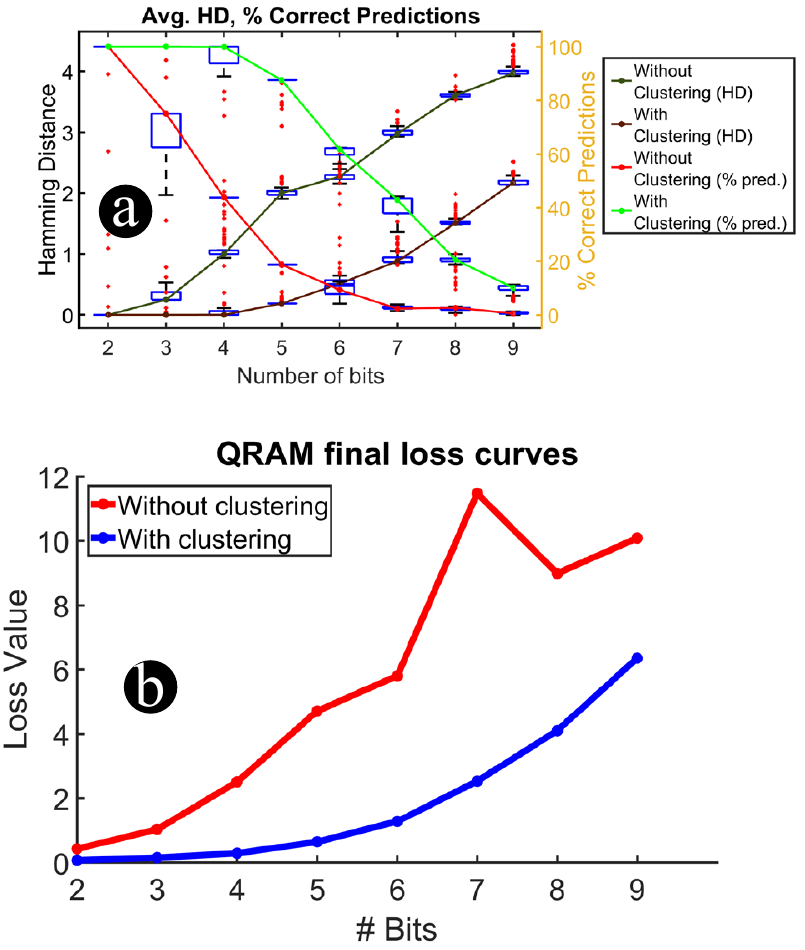}
    \vspace{-4.5mm}
    \caption{Binary QRAM results with and without clustering, (a) Hamming Distance (HD), percentage correct prediction and (b) final loss curve for different number of address lines.}
    \label{fig:binary_qram_graphs}
    \vspace{-5mm}
\end{figure}

From the results, it can be noted that, (i) the higher classification accuracy relative to EQGAN indicates that quantum \emph{read} operation performed by generatively learning a general data distribution is less accurate than mapping an address value to a digit image through training an extra auxiliary PQC network, (ii) the faster convergence relative to QNN with embedding indicates that QRAM storage realized by pre-training both QRAMs (main and auxiliary) together could pre-process the \emph{write} operation, thereby accelerating the classification task where fine-tuning is required. The pre-training phase relates more to \emph{write} to QRAM, whereas fine-tuning more to \emph{read} operation and, (iii) the faster convergence of two quantum algorithms relative to FCNN indicates potential quantum computing advantages with stronger representation powers of QNNs, as extensively demonstrated in the QNN community.

% \hl{[course materials should be avoided - suggest to remove this paragraph]} \hl{Traditionally, there are three main sources of error in machine learning: approximation error which is model dependent, estimation error which is data dependent, and optimization error which is algorithm dependent. For all the three cases, we are using the same data (digits dataset) and the same optimization method (Adam optimization). The key difference is in the selection of model. From the trends of the plots for all the three setups we can see that QRAM + QNN setup provides the fastest convergence, meaning that this model leads to low approximation error. In other words, this model is closest to the unknown mapping function $f$ that maps input address to the output data. This is the quantum advantage of QRAM storage provided by the QNN with QRAM embedding compared to the other two setups.}

\vspace{-0.4cm}

\section{QRAM for storage of binary data}

% \begin{figure}
%     \centering
%     \includegraphics[width=0.9\linewidth]{fig/CAL_Bin_loss.pdf}
%     \caption{Loss curve of binary data QRAM.}
%     \label{fig:binary_qram_loss}
% \end{figure}

\subsection{Experimental setup}
For storage of binary data, we feed address value as input and get data value as the output from the QRAM. 
%The number of address lines and data lines are kept the same i.e., the number of qubits used. We plot the HD between predicted output and correct output data, and also plot the percentage correct predictions with varying number of address lines.
The QRAM structure is defined based on the number of address lines and data lines used. 
For each setup, we keep the number of address lines and data lines equal. 
% okay-- lets discuss for date--\hl{[not necessarily equal - we can try larger number of address lines (larger means more features - digits accuracy is 100\% because of large features) to improve accuracy for the ASPLOS work]}. 
We send the binary address into quantum format and the QRAM gives approximate classical data as the output. In order to embed each address bit into a single qubit, we keep the number of qubits same as the number of address and data lines. %repeated--The PQC structure of the QRAM varies with number of address lines. 
%In Fig. \ref{fig:basic_qram_arch}, we keep circular layers of ansatz in Fig. \ref{fig:basic_qram_arch} followed by strongly entangling layers. 
For an $n$-address QRAM, the number of address-data pairs will be $2^n$, and for each address, the data is chosen randomly in the range of $\{0,1,2,....2^n-1\}$. $2^n$ address-data pairs are too low to train the QRAM when $n$ is low. In order to compensate for the low number of datapoints, we replicate the entire dataset of $2^n$ datapoints repeatedly 
%\hl{[Strictly speaking, this processing is problematic from machine learing perspective - we usually enlarge dataset by augmentation (add different but similar data points) rather than replication.]} 
till the dataset becomes large enough to train the QRAM parameters correctly. We call this dataset expansion. For example in a 2-address QRAM, if the address-data pairs are $\{00-01,01-11,10-00,11-01\}$, these 4 pairs are repeated continuously $\{00-01,01-11,10-00,11-01,00-01,01-11,10-00,11-01,...\}$ until the overall dataset size is large enough. Next, we train the QRAM PQC for 100 epochs and evaluate the average HD per epoch and calculate the percentage correct predictions per epoch. For any particular datapoint if the Hamming Distance (HD) between the predicted data output and actual data is zero we assume that the prediction is correct. If the HD is non-zero we assume that the prediction is incorrect. These metrics are evaluated for 100 epochs from 2 address lines to 9 address lines.
%For 100 epochs, these metrics are evaluated for a single address point. This process is repeated from 2 address lines to 9 address lines. %The data is presented in the form of box plots in Fig. \ref{fig:binary_qram_graphs}.
\vspace{-3mm}

\subsection{Experimental results}
From Fig. \ref{fig:binary_qram_graphs}, we note a general trend of increase in Hamming distance and decrease in percentage correct predictions with increase in number of address lines. This is expected since presence of more datapoints makes it harder for the QRAM to predict individual bits of these datapoints correctly. We get average Hamming distance of 0 and 100\% correct predictions only for a 2-address QRAM. After that, we start observing non-zero Hamming distance (HD = 0.25 for 3-address QRAM). There is also a drop in percentage correct predictions which varies from 75\%(3-address QRAM) to 0.58\%(9-address QRAM). The HD in this range is approximately 8\%(0.25, 3-address QRAM)-45\%(3.97, 9-address QRAM) of the total number of data lines meaning that on an average 8-45\% of the data bits are being mispredicted. We also show the loss curves in Fig. \ref{fig:binary_qram_graphs}(c). 
For every epoch mean squared error is calculated which increases with address lines.
% For every epoch, the loss calculated is the mean \color{blue}squared \color{black}error loss. It is observed that the final MSE loss increases as the number of address lines increase. 

\vspace{-3mm}
\subsection{Clustering based pre-processing and results}
We also propose an agglomerative clustering pre-processing step prior to storing the data in QRAM to improve the performance of the QRAM at higher number of address lines. The idea is to cluster the data and allocate one QRAM per cluster for better training. The clustering is performed in such a way that the average HD within each cluster is minimized. Since the data is divided into clusters, the number of individual datapoints per QRAM are lower, making the mapping from address to data easier.
%For each cluster, there is a separate QRAM PQC which stores address-data pairs in that cluster. 
With clustering, we observe zero HD and 100\% correct predictions upto 4-address QRAM and from 5-address QRAM to 9-address QRAM the HD increases from 0.18 to 2.17 and percentage correct predictions reduces from 87.5\% to 10.1\% (Fig. \ref{fig:binary_qram_graphs}). We also observe that the loss for each address line is lower with clustering as compared to the case without clustering.

%mostly repeated from above--\hl{Clustering data based on minimum average HD for every cluster makes the storage of binary data in the QRAM effective. Since HD is low, there will be lot of data bits that are identical, for lots of data within a cluster [could be rephrased here]. This will make the cluster QRAM easier to train the address-to-data mapping function. Moreover, since the data is divided into clusters, the number of individual datapoints per QRAM are lower, making the mapping from address to data easier.}
\vspace{-7mm}
\section{Conclusion}
\vspace{-2mm}
We presented QRAM models for ML task and binary data storage. For ML, we showed that QRAM + QNN provides faster convergence compared to FCNN and QNN with embedding, and the overall loss is lesser than FCNN, and slightly higher than QNN with embedding. For storage of binary data, we calculated HD and percentage correct predictions from 2 to 9 address lines and improved the results via an agglomerative clustering based pre-processing. In all the experiments performed, the PQC structure of the QRAM is kept fixed, and also the QRAM does not support storage of incremental data. A detailed analysis of the impact of the choice of PQC ansatz and task-incremental learning \cite{van2019three} of the QRAM PQC can be explored in the future.

% or


%\vspace{-0.4cm}
%\ifCLASSOPTIONcompsoc
%  \section*%{Acknowledgments}
%\else
%\fi

\textbf{Acknowledgments:} The work is supported by NSF (CNS -1722557, CNS-2129675, CCF-2210963, CCF-1718474, OIA-2040667, DGE-1723687, DGE-1821766, and DGE-2113839) and seed grants from Penn State ICDS.

%and Huck Institute of the Life Sciences. 
%We acknowledge the use of Pennylane services \cite{bergholm2018pennylane} for our research in quantum machine learning. 
%The views expressed are those of the authors, and do not reflect the official policy or position of Pennylane.

\ifCLASSOPTIONcaptionsoff
  \newpage
\fi



%\bibliographystyle{IEEEtran}
\vspace{-0.2cm}
\bibliographystyle{IEEEtran}
\bibliography{references}


% \end{IEEEbiography}


% % insert where needed to balance the two columns on the last page with
% %\newpage



%\vfill


\end{document}